# Deep multimodal saliency parcellation of cerebellar pathways: linking microstructure and individual function through explainable multitask learning

Ari Tchetchenian[1*], Leo Zekelman[2,3], Yuqian Chen[2], Jarrett Rushmore[4,5], Fan Zhang[6], Edward H. Yeterian[7], Nikos Makris[4,8], Yogesh Rathi[2,8], Erik Meijering[1], Yang Song[1*], Lauren J. O'Donnell[2]

[1] Biomedical Image Computing Group, School of Computer Science and Engineering, University of New South Wales (UNSW), Sydney, NSW, Australia.
[2] Department of Radiology, Brigham and Women's Hospital, Harvard Medical School, Boston, USA
[3] Harvard University, Cambridge, USA
[4] Departments of Psychiatry, Neurology and Radiology, Massachusetts General Hospital, Harvard Medical School, Boston, USA
[5] Department of Anatomy and Neurobiology, Boston University School of Medicine, Boston, USA
[6] School of Information and Communication Engineering, University of Electronic Science and Technology of China, Chengdu, China
[7] Department of Psychology, Colby College, Waterville, ME USA
[8] Department of Psychiatry, Brigham and Women's Hospital, Harvard Medical School, Boston, USA

*Correspondence: ari.tchetchenian@gmail.com (A.T.), yang.song1@unsw.edu.au (Y.S.)

**Data availability statement**
The Human Connectome Project minimally preprocessed young adult dataset imaging data and associated NIH toolbox measures are publicly available at https://db.humanconnectome.org/. The ORG tractography atlas is publicly available at http://dmri.slicer.org/atlases/ and code to apply the atlas is publicly available at https://github.com/SlicerDMRI/whitematteranalysis. All code developed for our experiments, as well our cerebellar parcellations will be publicly available at https://github.com/SlicerDMRI/DeepMSP.

**Acknowledgments**
This research was undertaken with the assistance of resources from the National Computational Infrastructure (NCI Australia), an NCRIS enabled capability supported by the Australian Government. This research has been supported by an Australian Government Research Training Program (RTP) Scholarship. Data were provided [in part] by the Human Connectome Project, WU-Minn Consortium (Principal Investigators: David Van Essen and Kamil Ugurbil; 1U54MH091657) funded by the 16 NIH Institutes and Centres that support the NIH Blueprint for Neuroscience Research; and by the McDonnell Center for Systems Neuroscience at Washington University. We gratefully acknowledge funding provided by the following National Institutes of Health (NIH) grants: R01MH132610, R01MH125860, R01MH119222, P41EB015902, R01NS125307, R01NS125781,

R01MH112748, R01MH111917, R01AG042512, R21DA042271, and K24MH116366. Fan Zhang is supported by the National Key R&D Program of China (No. 2023YFE0118600) and the National Natural Science Foundation of China (No. 62371107). Additionally, this study was made possible by the support of the UNSW-USA Networks of Excellence Grant.

**Conflict of interest**

None

**IRB statement**

The creation of the WU-Minn HCP dataset was approved by the institutional review board of Washington University in St. Louis (IRB #201204036)

# Abstract

**Introduction**: Parcellation of human cerebellar pathways is essential for advancing our understanding of the human brain. Existing diffusion MRI tractography parcellation methods have been successful in defining major cerebellar fibre tracts, while relying solely on fibre tract structure. However, each fibre tract may relay information related to multiple cognitive and motor functions of the cerebellum. Hence, it may be beneficial for parcellation to consider the potential importance of the fibre tracts for individual motor and cognitive functional performance measures.

**Methods**: In this work, we propose a multimodal data-driven method for cerebellar pathway parcellation, which incorporates both measures of microstructure and connectivity, and measures of individual functional performance. Our method involves first training a multitask deep network to predict various cognitive and motor measures from a set of fibre tract structural features. The importance of each structural feature for predicting each functional measure is then computed, resulting in a set of structure-function saliency values that are clustered to parcellate cerebellar pathways. We refer to our method as *Deep Multimodal Saliency Parcellation (DeepMSP)*, as it computes the saliency of structural measures for predicting cognitive and motor functional performance, with these saliencies being applied to the task of parcellation.

**Results**: Applying DeepMSP to a large-scale dataset from the Human Connectome Project Young Adult study (n=1,065), we found that it was feasible to identify multiple cerebellar pathway parcels with unique structure-function saliency patterns that were stable across training folds. We thoroughly experimented with all stages of the DeepMSP pipeline, including network selection, structure-function saliency representation, clustering algorithm, and cluster count. We found that a 1D convolutional neural network architecture and a transformer network architecture both performed comparably for the multitask prediction of endurance, strength, reading decoding, and vocabulary comprehension, with both architectures outperforming a fully connected network architecture. Quantitative experiments demonstrated that a proposed low-dimensional saliency representation with an explicit measure of motor vs cognitive category bias achieved the best parcellation results, while a parcel count of four was most successful according to standard cluster quality metrics. Our results suggested that motor and cognitive saliencies are distributed across the cerebellar white matter pathways. Inspection of the final k=4 parcellation revealed that the highest-saliency parcel was most salient for the prediction of both motor and cognitive performance scores and included parts of the middle and superior cerebellar peduncles.

**Conclusion**: Our proposed saliency-based parcellation framework, DeepMSP, enables multimodal, data-driven tractography parcellation. Through utilising both structural features and functional performance measures, this parcellation strategy may have the potential to enhance the study of structure-function relationships of the cerebellar pathways.

# 1 Introduction

The cerebellum is increasingly understood as both a motor and cognitive structure [Koziol et al., 2014]. Many parcellation approaches exist for the anatomy of the human cerebellum [Carass et al., 2018; Diedrichsen et al., 2009; Makris et al., 2005] and its cortical functional networks that often highlight the cerebellum's role in motor and cognitive performance [Buckner et al., 2011; Nettekoven et al., 2023]. However, parcellations of the cerebellar pathways as reconstructed by diffusion magnetic resonance imaging (dMRI) tractography have received less attention. dMRI tractography is the only method that allows in-vivo, non-invasive mapping of the human brain's white matter connections (Basser et al., 2000). Existing tractography-based parcellations define major cerebellar fibre tracts, including the inferior, middle, and superior peduncles [van Baarsen et al., 2016; Tang et al., 2018; Yeh et al., 2018; Zhang et al., 2018b], which can be subdivided more finely using fibre clustering [Zhang et al., 2018b] or according to cerebellar cortical terminations of streamlines [Rousseau et al., 2022]. These parcellations are powerful for enabling quantification of tractography [Zhang et al., 2022] to study the cerebellar pathways in health and disease [Beez et al., 2022; Filippi et al., 2018; Gupta et al., 2021; Habas and Manto, 2018; Harrison et al., 2021; Mittal et al., 2014; Phillips et al., 2015; Quartarone et al., 2020]. As the cerebellum contains 80% of the total number of neurons in the brain [Azevedo et al., 2009], and models of cerebellar pathways are helpful for the understanding of various conditions such as Parkinson's disease [Haghshomar et al., 2022], Alzheimer's disease [Toniolo et al., 2020], and autism spectrum disorder [Jeong et al., 2014], it can be clinically beneficial to develop and improve in vivo methods of cerebellar pathways parcellation.

However, current cerebellar pathway parcellation methods utilise fibre tract structure, but do not leverage the relationships between fibre tract structure and individual functional performance measures, such as motor or cognitive performance. In the field of brain parcellation, state-of-the-art methods frequently rely on multimodal sources of information, including anatomical, functional, connectional, cytoarchitectonic, and/or microstructural neuroimage data to define parcels within the brain's grey or white matter [Arslan et al., 2018; Eickhoff et al., 2015; Eickhoff et al., 2018; Glasser et al., 2016; Zhang et al., 2022]. In the cerebellum, functional information is known to be important for parcellating the cerebellar cortex [Nettekoven et al., 2023]. Recent work has demonstrated relationships between the structure of cerebellar pathways and individual cognitive performance in health and disease [Chang et al., 2022; Chen et al., 2020; Fritz et al., 2022; Kim et al., 2021; Zekelman et al., 2023]. Therefore, in this work we explore a parcellation method that can leverage the relationship between two sources of multimodal information, including individual structural (tractography) and functional (from individual functional performance measure data) information.

By training a network to predict functional performance measures from structural features, deep learning provides a data-driven approach that can encode this relationship between individual cerebellar pathway structure and individual functional performance. In the domain of medical imaging, there have been a variety of successful deep learning architectures. These range from convolutional neural networks [Sarvamangala and Kulkarni, 2022], to graph neural networks [Zhang et al., 2023] and transformers [Shamshad et al., 2023]. Despite being the most recent of these architectures, transformers have already achieved state-of-the-art results across a variety of imaging [Chen et al., 2021; Matsoukas et al., 2021; Usman et al., 2022] and non-imaging [Deng et al., 2022; Peng et al., 2021; Rao et al., 2022] tasks in the medical field. Transformers [Vaswani et al., 2017] utilise an attention mechanism that allows for better handling of long sequences of data by assigning importance to different sections of the input [Lin et al., 2022]. This makes transformers appealing for our task of neuroimaging-based individual functional performance prediction, where a large sequence of features (1,940 dMRI microstructural and connectivity measures) is input to the model. Additionally, the paradigm of multitask learning [Caruana, 1997] has gained traction as a model-agnostic method for enhancing the performance of deep models by having the model learn to predict multiple tasks simultaneously. Multitask learning enables a model to learn features that are relevant across the multiple tasks, reducing the chance of learning spurious or irrelevant features [Ruder, 2017], and has been shown to improve the performance of deep networks across many neuroimaging tasks [He et al., 2020; Liang et al., 2021; Liu et al., 2019].

To quantify the structure-function relationships learned by a deep model, we leverage the computer vision concept of *saliency*, which measures the importance of input features for a model's predictions. In a deep learning context, a saliency estimation method assigns a value to each input feature of the network, which represents the importance of that feature to the network's final prediction [Fan et al., 2021]. There exist a variety of saliency estimation methods, including gradient-based [Selvaraju et al., 2020; Simonyan et al., 2013; Springenberg et al., 2014], perturbation-based [Fong and Vedaldi, 2017; Ribeiro et al., 2016], and activation-based [Shrikumar et al., 06--11 Aug 2017; Zhou et al., 2016] methods. In this work, we use the gradient-based 'saliency maps' [Simonyan et al., 2013] approach, where saliency values for a particular target output are the partial derivative of the network's target output with respect to the input. In comparison to competing methods, the saliency maps approach has been shown to be invariant to constant shifts in input [Kindermans et al., 2019] and is sensitive to model parameters and output labels [Adebayo et al., 2018], both of which are desirable properties for saliencies.

Hence, in this work, we propose *Deep Multimodal Saliency Parcellation (DeepMSP)*, a multimodal data-driven approach for human cerebellar pathway parcellation, which leverages the relationship between the structural and functional information of individuals. We train a multitask learning transformer to take a set of high-dimensional dMRI measures describing the microstructure (fractional anisotropy, trace of tractography tensor) and connectivity (number of streamlines, number of points) of various fibre clusters, and predict performance on multiple individual functional performance measures from the NIH Toolbox. By investigating the resulting trained network, we can estimate the saliency of each dMRI feature and the fibre cluster from which it was measured, for the prediction of individual functional performance measures. We refer to this measure of importance as the *structure-function saliency*, as it encodes the importance of a structural measure for functional prediction. By then grouping fibre clusters with similar structure-function saliencies into parcels, we aim to discover meaningful parcellations of the cerebellar pathways. This overall strategy can enable the usage of high-dimensional, multimodal, structural and functional information to enhance the parcellation of white matter connections. Overall, in this work we obtain new insights into the predictive importance of various dMRI measures and the potential relevance of particular parcels within pathways for studying structure-function relationships of the human cerebellar pathways.

## 2 Materials & Methods

### 2.1 Overview

DeepMSP (Fig. 1) includes four major steps. First, input tractography data are processed to extract finely parcellated fibre clusters from which structural features are extracted to describe microstructure and connectivity. These parcels are based only on tractography streamline trajectory information and therefore have no inherent functional relevance, but can serve as a substrate to enable the definition of parcels with potential functional relevance. Second, the proposed multitask deep learning network is trained to predict individual functional performance measures from the structural features. Third, during inference on testing data, saliencies are measured to describe the impact of each structural feature from each fibre cluster on the prediction of individual functional performance measures. We call these *structure-function saliencies*. Fourth, saliency values are used as input features to a clustering algorithm, which discovers parcels with common saliency patterns. The final result is a parcellation of cerebellar pathways.

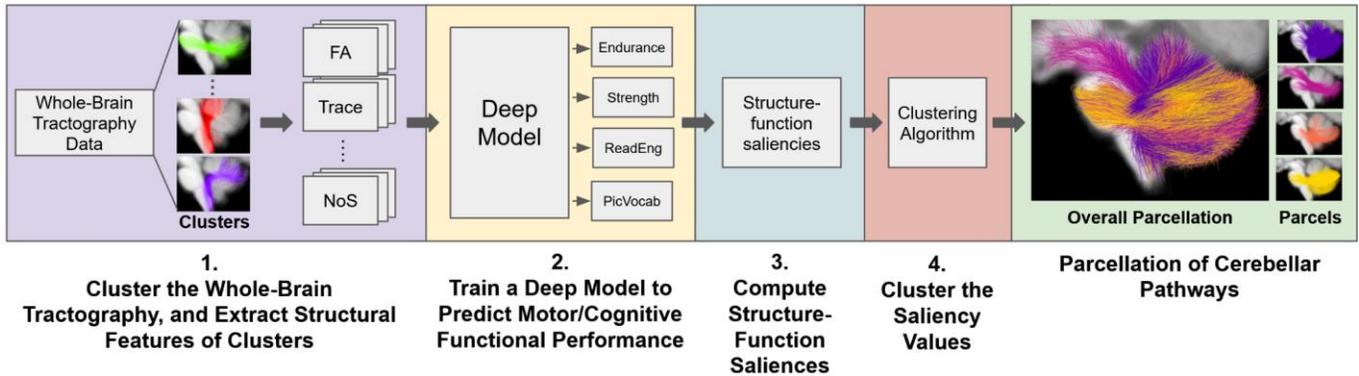

Fig 1: An overview of DeepMSP. 'FA' and 'NoS' refer to fractional anisotropy and number of streamlines. Endurance, Strength, ReadEng (reading decoding), and PicVocab (vocabulary comprehension) are examples of individual functional performance measures.

## 2.2 Dataset

We studied NIH Toolbox and dMRI data from the 1,065 subjects of the Human Connectome Project minimally preprocessed young adult dataset (HCP-YA) [Glasser et al., 2013]. The HCP-YA dataset consisted of 46% male and 54% female subjects, from 22 to 35 years old, where all dMRI scans were preprocessed according to the Human Connectome Project's minimal preprocessing pipeline. This pipeline included normalisation of b0 image intensity across runs, removal of various distortions (EPI distortions, eddy-current-induced distortions, subject motion), correction for gradient nonlinearities, registration of dMRI data to structural scans, bringing the data into 1.25mm structural space, and masking the dMRI data with a brain mask [Glasser et al., 2013]. Regarding the NIH Toolbox measures, we studied all age-adjusted motor and cognitive NIH Toolbox measures (Table 1) [Gershon et al., 2013; Weintraub et al., 2013]. We studied all measures due to the known importance of the cerebellum for both motor and cognitive processing [Schmahmann, 2019] and the recently reported relationships between cerebellar input and output pathways (peduncles) and individual performance across cognitive domains in many diseases [Chang et al., 2022; Chen et al., 2020; Fritz et al., 2022; Kim et al., 2021]. In healthy individuals, our group recently showed that cerebellar pathway microstructure significantly covaries with multiple NIH toolbox cognitive performance assessments in the HCP-YA dataset [Zekelman et al., 2023].

Saliency-based Cerebellar Parcellation

| Category | NIH Toolbox Measure | Description |
|---|---|---|
| Motor | Endurance | Measures cardiovascular endurance by the distance a participant walks in 2 minutes on a 50-foot course. |
| | Locomotion (GaitSpeed) | Assesses walking speed by timing how quickly a participant walks a 4-metre distance at their usual pace. |
| | Dexterity | Evaluates manual dexterity by timing how quickly a participant can insert and remove 9 pegs into a pegboard with their dominant hand. |
| | Strength | Measures grip strength by having participants squeeze a Jamar Plus Digital dynamometer, which provides a digital reading of force in pounds. |
| Cognitive | Episodic memory (PicSeq) | Measures memory retention through the recall of a sequence of pictures shown on a computer screen. |
| | Cognitive flexibility (CardSort) | Assesses the ability to switch between tasks by asking participants to match pictures based on varying dimensions like shape and colour. |
| | Inhibition (Flanker) | Measures attention and inhibitory control by requiring focus on a central stimulus while ignoring surrounding stimuli. |
| | Reading decoding (ReadEng) | Evaluates reading skills in English or Spanish by asking participants to read and pronounce letters and words accurately. |
| | Vocabulary comprehension (PicVocab) | Tests understanding of vocabulary by presenting a spoken word and asking participants to select the corresponding picture. |
| | Processing speed (ProcSpeed) | Measures the speed at which a participant can determine if two side-by-side pictures are the same or different. |
| | Working memory (ListSort) | Assesses the ability to remember and sequence visually- and orally-presented items, such as foods and animals, in specific conditions. |

Table 1: A description for each NIH toolbox motor and cognitive measure used in this paper. Abbreviations for NIH Toolbox measures are in brackets.

The HCP-YA dMRI data were processed to perform whole brain tractography in each subject using the two-tensor unscented Kalman filter (UKF) method [Reddy and Rathi, 2016] as implemented in the ukftractography package (https://github.com/pnlbwh/ukftractography) following published methods [He et al., 2022; Zekelman et al., 2022]. In contrast to other tractography methods that fit a model to the diffusion signal independently at each voxel, UKF employs prior information from the previous tracking step to stabilise model fitting. The two-tensor model can reconstruct fibre crossings that are prevalent in the brain and brainstem [Ford et al., 2013; Jeurissen et al., 2013]. The first tensor is associated with the tract being traced, while the second tensor models fibres that cross through the tract. For each subject, the whole brain tractogram was divided into fibre clusters with a robust machine-learning approach that consistently extracts white matter connections across datasets, acquisitions, and the human lifespan [Zhang et al., 2018b] as implemented in the WMA package (https://github.com/SlicerDMRI/whitematteranalysis). This fibre clustering approach has been previously demonstrated to be successful for tractography approaches beyond the UKF method used in this paper, for instance diffusion tensor and constrained spherical deconvolution tractography [Zhang et al., 2018b]. The approach first involves the registration of the subject's tractogram to an atlas tractogram [O'Donnell et al., 2012], followed by a spectral embedding to represent each streamline using its similarity to thousands of streamlines in the atlas [O'Donnell and Westin, 2007]. Next, spectral embedding clustering is performed to extract subject-specific anatomical fibre clusters according to the atlas. We used an anatomically curated white matter tract atlas, the ORG tractography atlas [Zhang et al., 2018b] (http://dmri.slicer.org/atlases/) provided by SlicerDMRI [Norton et al., 2017; Zhang et al., 2020]. The atlas defines 97 cerebellum fibre clusters, which are anatomically categorised into five tract categories, including the inferior cerebellar peduncle (ICP), the middle cerebellar peduncle (MCP), the superior cerebellar peduncle (SCP), the input and Purkinje cell fibres (IP, including streamlines in the medullary white matter core of the cerebellum), and the putative parallel fibres (PF, including intra-cortical streamlines in the cerebellar cortex). We note that these fibre clusters describe a fine parcellation of the cerebellar pathways. In this work, our goal is to go beyond this structural parcellation to investigate parcellations that have the potential to be more meaningful for studying structure-function relationships of the cerebellar pathways. We used these input fibre clusters as a fine parcellation to enable the initial measurement of quantitative dMRI features in all subjects.

For a given subject, from the fibre clusters extracted using the previously described spectral embedding approach, we further extracted 20 quantitative dMRI structural features from the subset of fibre clusters belonging to the cerebellum. These features included microstructural and connectional geometric information measured within each fibre cluster. The microstructural features computed from the multi-tensor model included the minimum, maximum, median, mean, and variance of the fractional anisotropy of tensors 1 (FA1) and 2 (FA2), and the minimum, maximum, median, and mean trace of tensors 1 (Trace1) and 2 (Trace2). FA is a measure of the amount of directionality of the diffusion, while trace is a mathematical property derived from a diffusion tensor that measures the total amount of diffusion in a voxel [O'Donnell and Westin, 2011]. Rather than focusing on the traditional mean value [Zhang et al., 2022], employing multiple summary statistics to describe tissue microstructure in this way can benefit machine-learning applications [He et al., 2022; Zhang et al., 2018a]. In addition to these microstructural features, connectional geometric features described the "connectivity" and geometry of each cluster and included the number of streamlines (NoS) and the number of points (NoP) along streamlines. We note that while the NoS is popularly considered to be a measure of "connectivity" and is used in a large number of studies, it is only indirectly related to structural connectivity "strength" [Zhang et al., 2022]. Thus NoS and NoP can also be considered to be measures of the geometry of connections. In total, 20 dMRI features were extracted within each of the 97 fibre clusters, resulting in 1,940 input features per subject. Each input feature was standardised by subtracting the training set mean and dividing by the training set standard deviation before being fed into the model.

## 2.3 Model architecture

In order to perform parcellation, we required a network that was trained to predict individual functional performance measures so that we could investigate the fibre clusters that were most salient for its predictions.

We trained three multitask learning regression neural networks, each designed to predict 11 continuous individual functional performance measures from the NIH Toolbox, from an input of a 1D vector with 1,940 dMRI features. A 1D vector representation was chosen, as it was shown to outperform a 2D representation [He et al., 2022]. We chose to force the network to predict multiple NIH Toolbox measures simultaneously (i.e. multitask learning), as multitask learning has been demonstrated to improve network performance [He et al., 2020; Liang et al., 2021; Liu et al., 2019] by learning shared feature representations for all tasks that reduces the chance of learning irrelevant features correlated with only a single task [Ruder, 2017].

Our first architecture was a fully connected model comprised of a sequence of fully connected layers. This is the simplest possible deep network architecture where each neuron in one layer is connected to every neuron in the next layer (supplementary Fig. S1), and was chosen to establish a baseline. The second architecture was a 1D convolutional neural network (1DCNN) (supplementary Fig. S2), which was found to perform well in [He et al., 2022] for age and sex prediction. A 1DCNN utilises specialised layers called 1D convolutional layers, which integrate information from adjacent input elements via a learned convolution operation. The final model was a multi-head attention transformer [Vaswani et al., 2017] (Fig. 2), which excels in capturing long-range dependencies in the data by focusing attention on different regions of input [Lin et al., 2022]. Each model was balanced to contain approximately 20 million parameters, ensuring a fair comparison across architectures by isolating the effectiveness of the architecture design, independent of model capacity.

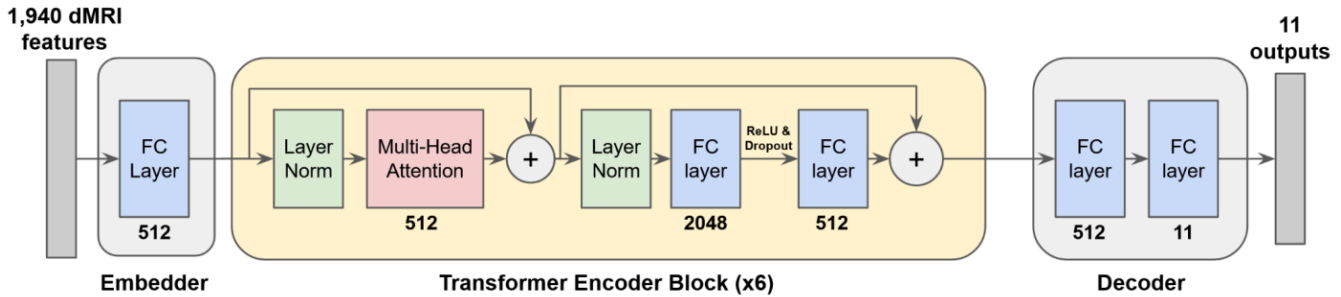

Fig 2: Architecture diagram of our multi-head attention transformer based on [Vaswani et al., 2017]. 'FC layer' indicates a fully connected layer, and the number under each layer indicates the output size of that layer. Multi-head attention used 8 heads, and the entire transformer encoder block was repeated 6 times. The output of the penultimate fully connected layer in each encoder block is followed by a ReLU activation function and a dropout layer. Note that layer normalisation is placed before the multi-head attention and fully connected layers rather than after, as this was shown to improve model training time in [Xiong et al., 13--18 Jul 2020].

## 2.4 Model training and evaluation details

All networks were trained using 5-fold cross-validation with a 60/20/20 train/validation/test split of 639/213/213 subjects, where each subject was randomly allocated to a single fold. We utilised a mean squared error (MSE) loss function, which was optimised over 50 epochs via the AdamW optimiser [Loshchilov and Hutter, 2017], with $\beta_1 = 0.9$, $\beta_2 = 0.999$, and a weight decay of 0.01. The learning rate was reduced by a factor of 0.1 when the validation loss had not decreased for 15 epochs. All models were implemented and trained using PyTorch on a National Computational Infrastructure (NCI Australia) node containing 16 GB of memory, 12 24-core Intel Xeon Cascade Lake processors, and an Nvidia V100 GPU with 32 GB of memory.

Each model's hyperparameters were optimised via a 50-epoch grid search on a single fold of cross-validation (learning rate [1e-3, 1e-4, 1e-5], batch size [1, 5, 10, 20], and dropout percentage [0, 10%, 50%, 90%]). The set of possible values for each hyperparameter was chosen to cover a broad range of values, while being limited to 3 values to allow for the completion of the grid search within a reasonable timeframe. Additionally, the upper end of the batch size values (20) was chosen based on GPU memory limits. The hyperparameter values resulting in the lowest validation set loss across the 50 epochs were as follows: the fully connected network had batch size 20, learning rate 1e-3, and no dropout; the 1DCNN had batch size 1, learning rate 1e-3, and no dropout; and the transformer had batch size 10, learning rate 1e-4, and 10% dropout. Models with these hyperparameters then underwent proper 5-fold cross-validation training and evaluation.

The prediction performance of each deep model was evaluated using two metrics: mean absolute error (MAE) and Pearson correlation coefficient (*r*) [Freedman et al., 2007]. MAE calculates the average magnitude of errors between predicted ($y_i$) and true ($x_i$) values, with lower values indicating better performance:

$$MAE = \frac{\sum_{i=1}^{n}|y_i - x_i|}{n} \quad (1)$$

The Pearson correlation coefficient measures the linear correlation between the predicted ($y_i$) and true ($x_i$) values, where values closer to 1 and -1 indicate strong positive and negative correlations respectively, while a value of 0 indicates no correlation. It is computed as follows, where $\bar{y}$ and $\bar{x}$ indicate the mean of the predicted and true values:

$$r = \frac{\sum (x_i - \bar{x})(y_i - \bar{y})}{\sqrt{\sum (x_i - \bar{x})^2 \sum (y_i - \bar{y})^2}} \quad (2)$$

The mean and standard deviation for these two metrics across all cross-validation test sets were reported for each NIH Toolbox measure independently. NIH Toolbox measures for which no model achieved $r \geq 0.1$ after subtracting the standard deviation, were determined to have performed insufficiently, and were discarded from further analysis [Gong et al., 2021]. The model that performed the best for the remaining NIH Toolbox measures was then selected for saliency computation and parcellation.

The statistical significance of performance differences for Pearson correlation coefficient values between models was determined via a repeated measures ANOVA, followed by post-hoc pairwise Tukey tests. The normality of residuals was verified using the Shapiro-Wilk test, and homogeneity of variances was assessed via Levene's test. Pairwise comparisons were deemed statistically significant if the post-hoc Tukey test resulted in a p-value below 0.05. The statistical significance of performance differences for MAE between models was determined as follows. Because the Shapiro-Wilk test indicated non-normality of MAE residuals, for MAE value comparisons we applied the non-parametric Kruskal-Wallis test to determine the statistical significance of MAE value differences, followed by a pairwise post-hoc Dunn's test with Bonferroni correction for multiple comparisons.

## 2.5 Computing structure-function saliencies as feature vectors for parcellation

A set of structure-function saliency values were generated for each NIH Toolbox measure that the network was trained to predict. This resulted in a 1,940-element vector (97 fibre clusters times 20 dMRI features per cluster) for each task of each subject, corresponding to the importance of each dMRI feature for the network's output prediction. To enable symmetric parcellation of cerebellar structures [Nettekoven et al., 2023], the saliency values of bilaterally defined fibre clusters [Zhang et al., 2018b] were averaged, resulting in a 1,060 element structure-function saliency vector, corresponding to 53 bilaterally defined fibre clusters. Symmetric averaging of saliency vectors across hemispheres can enhance stability to benefit this initial investigation. A benefit of symmetric tractography parcellation is that it provides corresponding structures across hemispheres, thus enabling the study of the structural and tissue microstructure lateralization of brain pathways [O'Donnell and Westin, 2007; Propper et al., 2010].

These saliency values were computed by taking the absolute value of task-specific saliency maps [Simonyan et al., 2013]. Saliency maps assign a value to each element of a deep network's input vector, which is the partial derivative of the network's output with respect to the input at the particular point of the current input vector. Practically, this is achieved by forward propagating the input data, resulting in an output vector. The output for the target class is then backpropagated to the input layer. Each input element is thus assigned a signed saliency value, the magnitude of the value being proportional to the amount of impact the corresponding input element has on the network's prediction for the target class, while the sign of the value indicates whether the element contributes to increasing (positive sign) or decreasing (negative sign) the network's predicted value. In our experimentation, we are explicitly interested in the magnitude of the importance of each feature for predicting NIH Toolbox measures, regardless of the direction of influence on the predicted measures, hence we disregard the sign of the saliency vectors by taking the absolute value.

When comparing saliency values across multiple NIH Toolbox measures, it is necessary to establish a consistent scale for valid comparison. Saliency values for different NIH Toolbox measures are produced at different scales due to being a gradient value tied to the scale of the NIH Toolbox measure. Hence, comparing the raw saliency values across different NIH Toolbox measures would be unproductive, as a particular feature's saliency value being larger or smaller between different measures could be due to it being more or less salient, but may also be due to the differing scales of the output spaces. We address this issue by applying min-max normalisation to each generated saliency vector, resulting in all elements falling within the range [0,1], representing the relative saliency of input features.

Computing the saliency values for each NIH Toolbox task, and averaging across all subjects to get a single model-level representation of structure-function saliency, results in a 1,060-element vector. This vector can be separated into a 20-element saliency vector for each of the 53 fibre clusters, where each value of the 20-element vector corresponds to one of the 20 dMRI features per fibre cluster. The simplest method of parcellation involves concatenating these 20-element vectors for each NIH Toolbox measure into a single 1D vector with N*20 elements, where N is the number of NIH Toolbox measures that have been deemed as sufficiently predictive (r >= 0.1 after subtracting standard deviation) by the deep model. However, this is a very high-dimensional vector, which is not ideal, as clustering algorithms are known to struggle in high-dimensional spaces [Domingos, 2012]. Hence, we experimented with a variety of dimensionality reduction approaches, including both automated approaches like principal component analysis (PCA), which reduces dimensionality by transforming the original elements of the vector into a new set of uncorrelated elements called principal components, and hand-crafted approaches (Table 2).

Saliency-based Cerebellar Parcellation| Name | Description | Dimensionality |
|---|---|---|
| Original Vector | The original saliency vector, which is a concatenation of the 20-element saliency vectors for each of the *N* NIH Toolbox measures. | 20*_N_ |
| PCA-8 | Applying PCA with 8 components to the original vector. | 8 |
| PCA-4 | Applying PCA with 4 components to the original vector. | 4 |
| Mean of NIH measures | Taking the average across the 20 dMRI saliencies for each NIH Toolbox measure to get a single average saliency value per NIH Toolbox measure. | 4 |
| Mean of Categories | As per 'Mean of NIH measures', but further averaging saliencies for all motor and all cognitive measures to get a single average saliency value for motor and cognitive categories. | 2 |
| Category Displacements | As per 'Mean of Category Saliencies' approach, but adding an element that explicitly measures the amount of bias toward one of the categories, calculated by: $$\frac{motor - cognitive}{\sqrt{2}}$$ This value is the signed perpendicular distance of a given <motor, cognitive> point from the motor = cognitive line in 2D space. | 3 |

Table 2: Descriptions of various saliency vector representations we use in our experiments. The dimensionality column indicates the number of elements in the saliency vector representing a fibre cluster when performing parcellation in Section 2.6.

## 2.6 Parcellating pathways using saliencies

Parcellation of the cerebellar pathways was achieved by applying clustering algorithms to the set of structure-function saliency vectors described in Section 2.5.

To choose a suitable clustering algorithm, we experimented with both k-means and agglomerative hierarchical clustering to parcellate the cerebellar pathways. These algorithms were chosen as they represent different clustering paradigms [Hastie et al., 2009]: partitioning (k-means) and hierarchical (agglomerative) [Rokach, 2010]. In k-means [MacQueen, 1967] clusters are formed by initialising *k* centroids, assigning clusters to their nearest centroid, and iteratively updating datapoint-centroid assignments until cluster centroids no longer significantly change. Alternatively, agglomerative hierarchical clustering [Hastie et al., 2009] is a bottom-up hierarchical clustering approach that iteratively merges nearest-neighbour datapoints into clusters until only a single cluster remains. The resulting dendrogram is then cut off at the desired height, where the number of branches at that height indicates the number of clusters.

Evaluation of brain parcellation is a known challenge: selection of a parcellation depends on the task at hand [Arslan et al., 2018] and there may not be an optimal number of parcels because of the multiscale nature of the brain [Eickhoff et al., 2018]. Therefore, we experimented with a [2,14] range for parcel counts, choosing values within that range for further downstream analysis based on performance across a variety of metrics. Since we did not have a ground-truth clustering dataset, we employed the following internal cluster quality metrics that evaluated the intrinsic properties of the clustering such as within-cluster dispersion and between-cluster distance: Silhouette coefficient (SC) [Rousseeuw, 1987], Davies-Bouldin index (DBI) [Davies and Bouldin, 1979]. These metrics are often applied to assess the quality of tractography parcellation results [Siless et al., 2013; Vázquez et al., 2020]. See supplementary Table S1 for the full descriptions and equations for these metrics. For interpretation purposes, we note that in traditional cluster analysis an SC >= 0.5 indicates a high-quality clustering result [Mooi and Sarstedt, 2011], while for the challenging problem of tractography parcellation, SC values can range from 0.2 to 0.4 [Siless et al., 2013]. DBI has also been used to evaluate tractography parcellation, with values around 0.7 indicating success (lower is better) [Vázquez et al., 2020].

# 3 Results

## 3.1 Choosing a deep network architecture

NIH Toolbox measures with satisfactory prediction performance ($r >= 0.1$ as in [Gong et al., 2021], described in section 2.4) included two motor (endurance, strength) and two cognitive (reading decoding, vocabulary comprehension) measures. For each of these measures, both the transformer and 1DCNN outperformed the fully connected network in terms of both Pearson's $r$ and MAE ($p < 0.05$) (Table 3), with the exception of the MAE of the strength measure, where there was no statistically significant difference between the fully connected network and the 1DCNN ($p = 0.071$). Additionally, there was no statistically significant difference in Pearon $r$ or MAE between the 1DCNN and transformer for each NIH toolbox measure. However, due to the mean scores for both Pearson $r$ and MAE of the transformer being slightly better than the 1DCNN for the vast majority of metrics (Table 3), as well as the previously stated lack of statistically significant difference between the fully connected network and 1DCNN for the strength measure, we chose to proceed with the transformer model for subsequent experiments. For a complete report of the performance of all networks for all 11 NIH Toolbox measures, see supplementary Table S2 and S3, and for a complete list of statistical p-values, see supplementary table S4.

| Category | NIH Toolbox Measure | Pearson's $r$ | | | MAE | | |
|---|---|---|---|---|---|---|---|
| | | Fully Connected | 1DCNN | Transformer | Fully Connected | 1DCNN | Transformer |
| Motor | Endurance | 0.09 (± 0.07) | 0.23 (± 0.06) | **0.25 (± 0.02)** | 17.89 (± 0.41) | **11.24 (± 0.76)** | 11.34 (± 0.76) |
| | Strength | 0.22 (± 0.17) | **0.57 (± 0.04)** | **0.57 (± 0.05)** | 19.88 (± 1.36) | 14.51 (± 1.34) | **13.51 (± 1.25)** |
| Cognitive | ReadEng | 0.02 (± 0.07) | 0.20 (± 0.05) | **0.21 (± 0.05)** | 18.64 (± 0.89) | 12.49 (± 1.01) | **12.38 (± 0.92)** |
| | PicVocab | 0.02 (± 0.06) | 0.21 (± 0.07) | **0.25 (± 0.07)** | 18.70 (± 0.55) | 12.01 (± 0.75) | **11.81 (± 0.94)** |

Table 3. The Pearson correlation coefficients and mean absolute errors (MAE) of the fully connected, 1DCNN, and transformer models for endurance, strength, reading decoding (ReadEng), and vocabulary comprehension (PicVocab) NIH Toolbox measures. Values are the mean across the 5 folds of cross-validation, with the standard deviation indicated in brackets. The results with the highest means for each NIH toolbox measure are bolded.

## 3.2 Structure-function saliency values and consistency across folds

Examining the saliency values computed from the multitask transformer (Fig. 3a), we found that the overall mean saliency for each dMRI feature category was relatively similar (0.24 for FA1, NoS, Trace1, and 0.25 for FA2, NoP, and Trace2). We also found that the 'min' statistic was the most salient regardless of feature category (mean saliency of 0.27), with more typical statistics used in dMRI prediction tasks: 'mean' and 'median', being the lowest 2 salient statistics for all categories (both with mean saliencies of 0.22). We also observe different patterns of saliencies across clusters when averaging cross motor and cognitive NIH Toolbox measures (Fig. 3b), with differences in motor and cognitive scores being larger toward the right of Fig. 3b. Since we have sorted clusters based on their mean motor and cognitive saliency, this indicates that the majority of the difference in saliency across motor/cognitive categories stems from the most salient clusters.

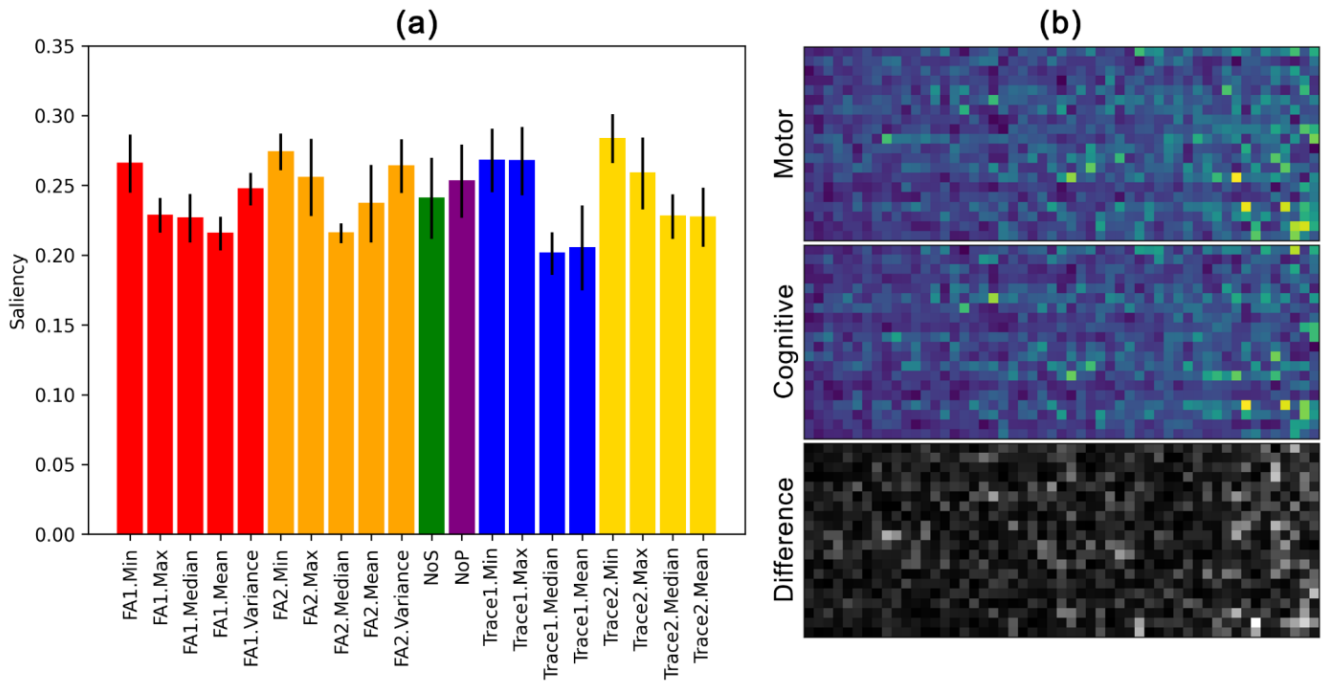

Fig 3: Visualising mean saliencies. Depicted in (a) is the mean of each dMRI feature, where colour indicates a common dMRI measure, and error bars indicate the standard deviation over folds of cross validation. Depicted in (b) is the mean motor, cognitive, and absolute difference between motor and cognitive saliencies, where each row corresponds to a dMRI feature, and each column corresponds to a fibre cluster. Results are averaged across all 1,065 subjects. Columns are sorted by the mean motor/cognitive score of the column.

Averaging the 1,060-element vectors across all subjects within a fold, we found very similar standard deviations across folds for each NIH Toolbox measure: 0.14 for strength, and 0.13 for endurance, reading decoding (ReadEng), and vocabulary comprehension (PicVocab). We also found that within each task, the 1,060-element saliency vectors were very similar, achieving mean pairwise cosine similarity scores of: 0.96 (± 0.00) for strength, and 0.93 (± 0.01) for endurance, ReadEng, and PicVocab.

## 3.3 Selecting a saliency vector representation

We found (Fig. 4) that reducing the dimensionality of the vectors that are being used for clustering results in higher silhouette coefficient values across a variety of cluster counts, for both k-means and agglomerative clustering. We selected the 'Category Displacements (3 elements)' representation for further analysis, as it exceeded the desired silhouette coefficient threshold of 0.5 [Mooi and Sarstedt, 2011] for both clustering methods, and achieved a higher median SC value (0.50 k-means, 0.49 agglomerative) compared to the 'Mean of Categories (2 elements)' representation (0.46 k-means, 0.44 agglomerative) which also exceeded the threshold.

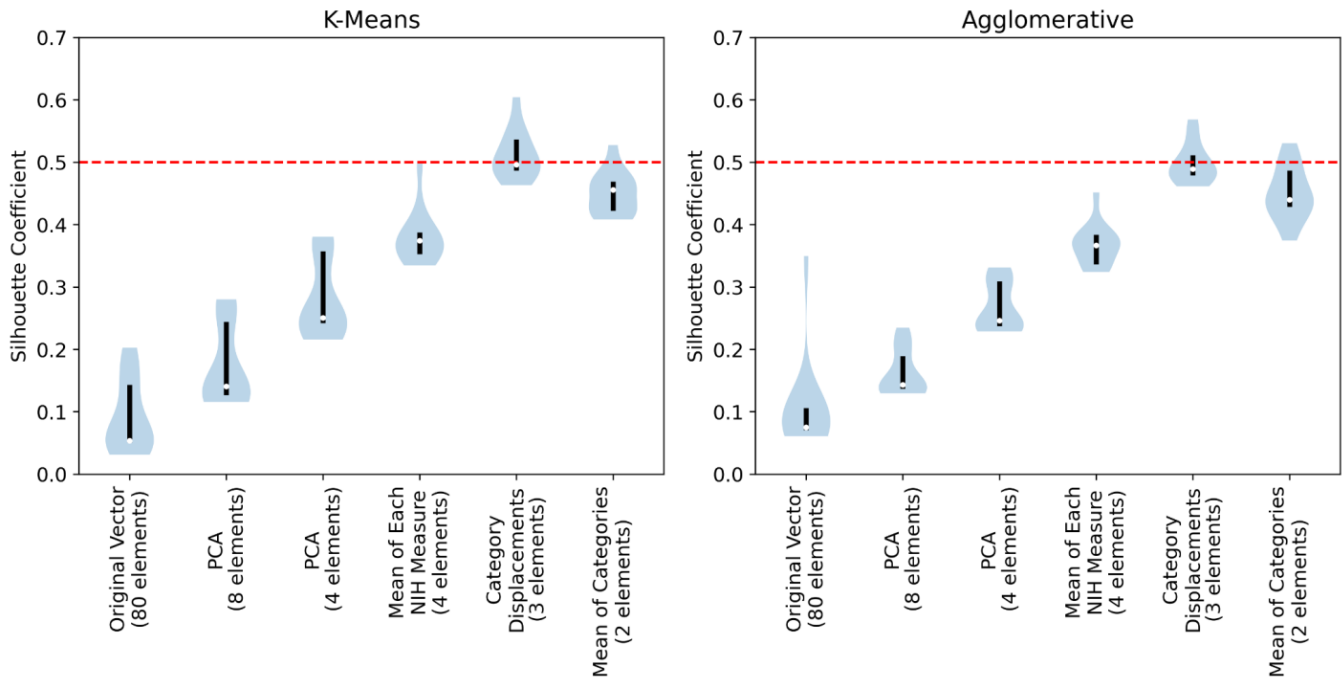

Fig 4: Violin plots of the silhouette coefficient values for k-means and agglomerative clustering, where clustering was applied to each vector representation across cluster counts in the range [2,14]. Black bars indicate interquartile range, with the white dot indicating the median value. A red horizontal line is set at a silhouette coefficient value of 0.5, which is a standard criterion for a good quality SC result [Mooi and Sarstedt, 2011].

## 3.4 Selecting the number of parcels and clustering algorithm

Investigating the 3-element category displacement feature representation more closely (Fig. 5), we found that a parcel count of 4 achieved the best internal cluster quality scores (k-means: 0.60 SC, 0.45 DBI; agglomerative: 0.57 SC, 0.46 DBI). Hence, we selected a parcel count of 4 for parcellating the cerebellar pathways. In regard to choosing a clustering algorithm, there were very small differences in performance between k-means and agglomerative clustering for a parcel count of 4, with k-means achieving equal or slightly better (5.7% higher SC, 2.17% lower DBI) performance across every metric. Hence, k-means was selected as the clustering algorithm for creating the cerebellar pathway parcellation.

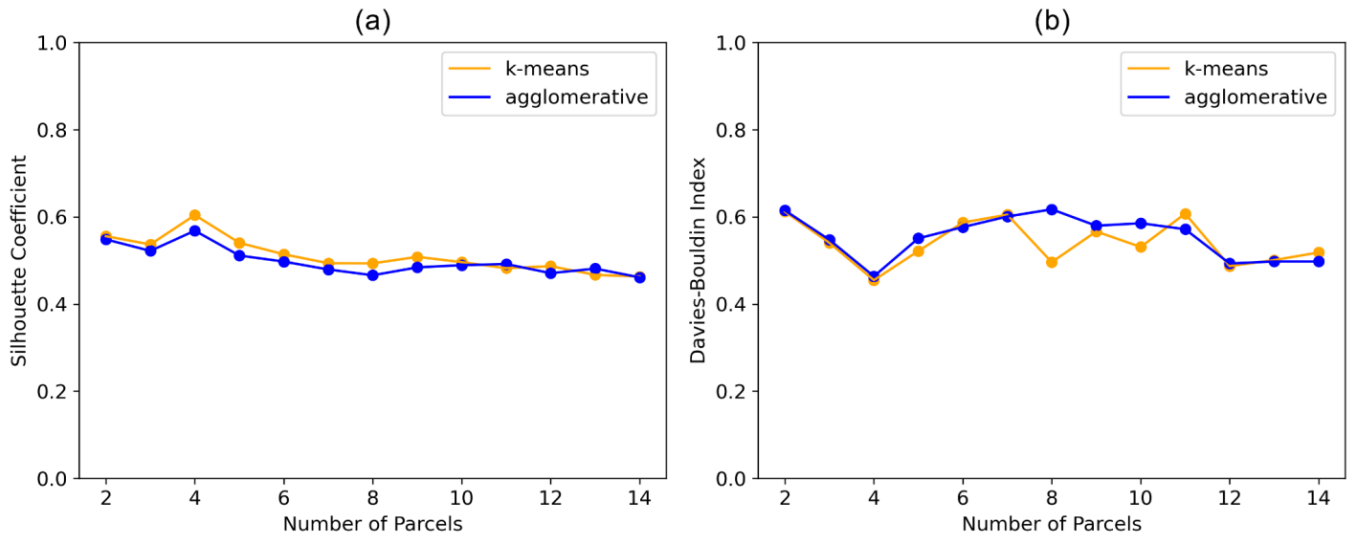

Fig 5: Clustering quality for parcel counts from 2 to 14. (a) Silhouette coefficient (higher is better), (b) Davies-Bouldin index (lower is better). Clustering was performed on 'Category Displacements' saliency vectors.

## 3.5 Parcellation

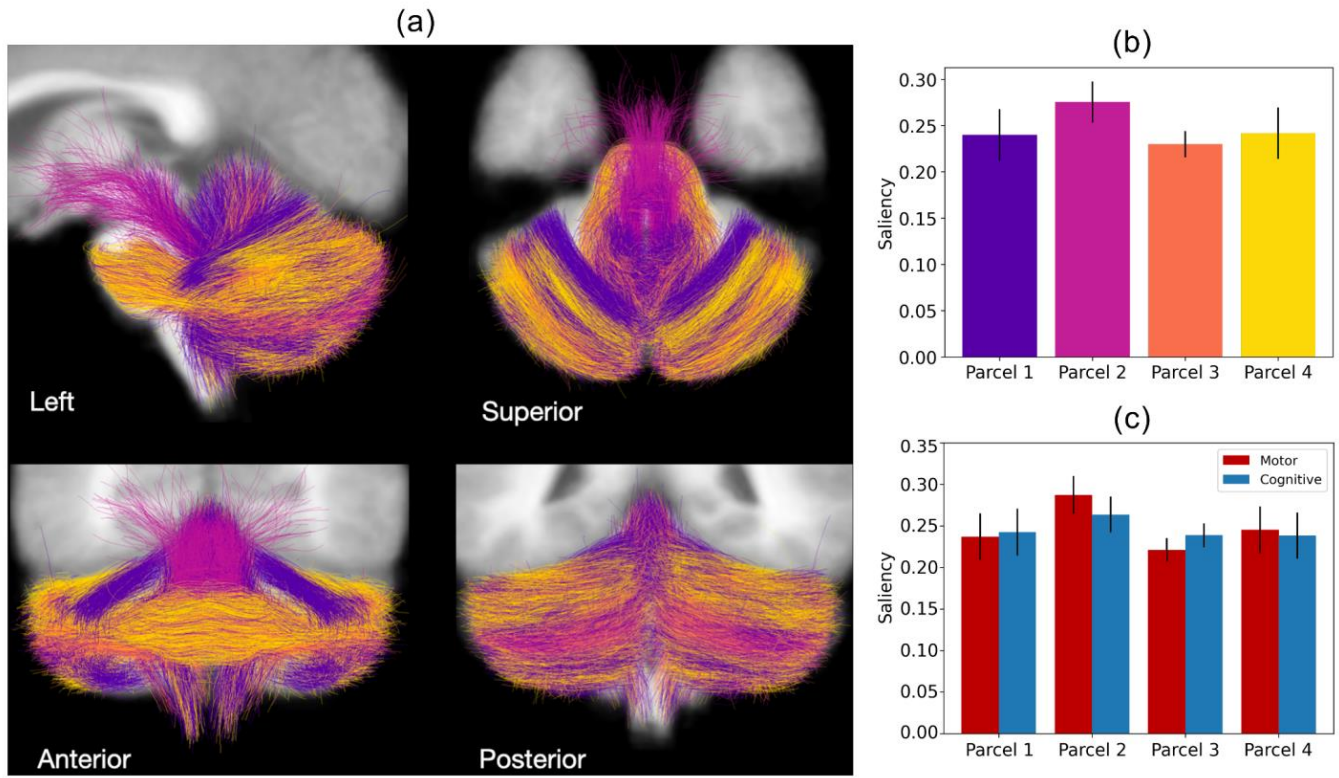

Fig 6: Visualisation of the cerebellar pathway parcellation. Shown in (a) is the overall parcellation overlaid on a T1-weighted MRI scan, with each parcel being a different colour, (b) is the mean saliency for each parcel, with error bars indicating standard deviation across clusters within each parcel, and the colour of each bar corresponding to the parcel colour in (a), (c) is the mean motor and cognitive saliency of each parcel.

Cerebellar fibre clusters were parcellated into 4 parcels (Fig. 6a). Parcel 2 was found to have the highest mean saliency (Fig. 6b) and was motor dominant (Fig. 6c), while parcel 3 was cognitive dominant, and parcels 1 and 4 had very similar motor and cognitive saliencies. See supplementary Fig. S3 for a more detailed breakdown of the mean saliencies for NIH Toolbox measures and dMRI features. Each parcel was also visually distinct (Fig. 7), had a different distribution of white matter tracts, and connected different cerebellar regions (Table 4, Fig. 8). As a complement to the preceding visualisations illustrating the parcels as structural connections, we include a visualisation of the parcels coloured by their relative motor vs cognitive saliency values (Fig. 9).

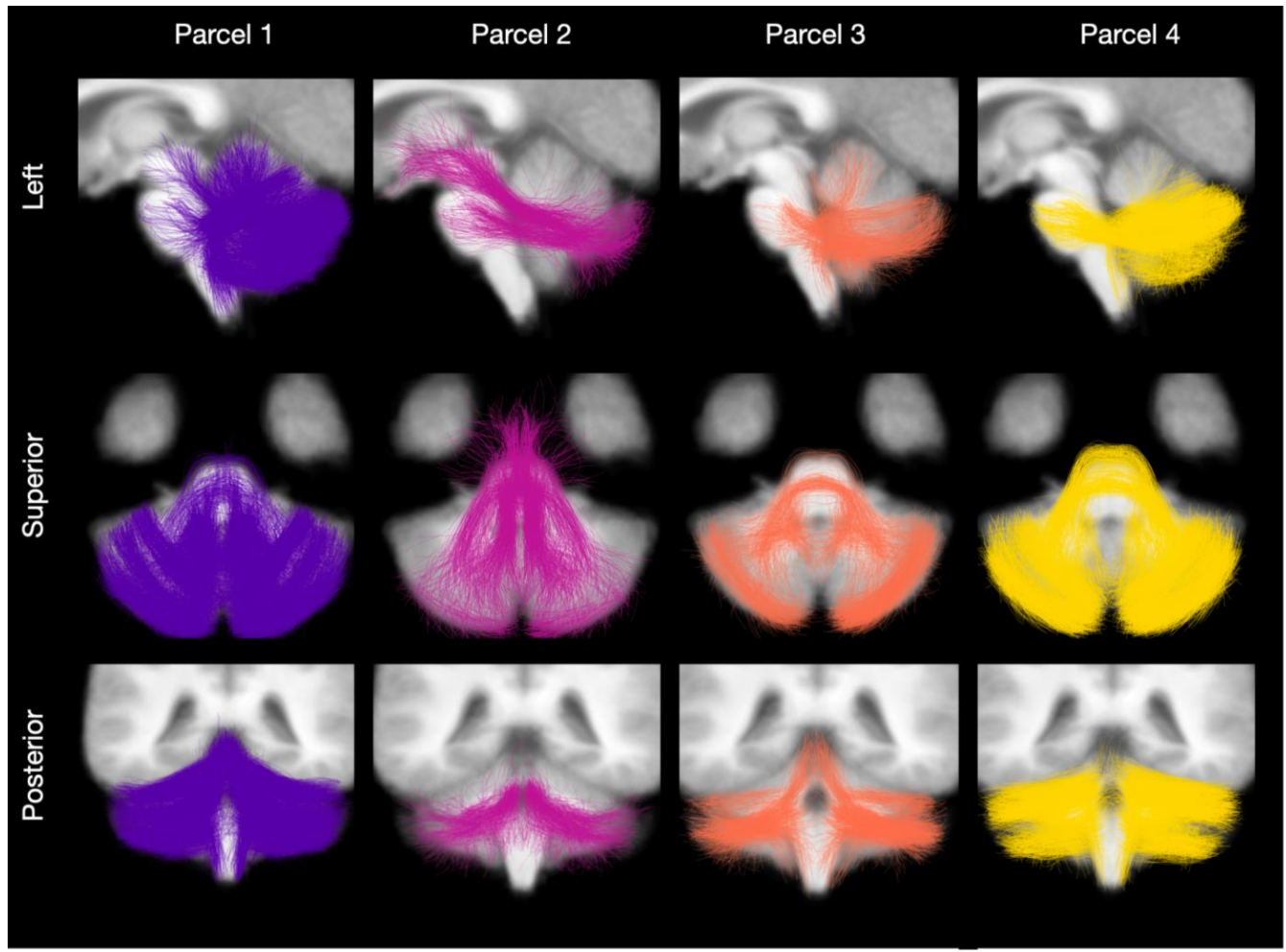

Fig 7: Visualisations of each saliency-based cerebellar pathway parcel overlaid on a reference T1-weighted MRI scan.

| Parcel | Number of Fibre Clusters in Each Tract | | | | | Top 5 Cerebellar Regions Connected |
|---|---|---|---|---|---|---|
| | ICP | MCP | SCP | IP | PF | |
| 1 | 2 | 1 | 1 | 8 | 10 | Crus II, Crus I, Lobule I_IV, Lobule VI, Dentate |
| 2 | 0 | 2 | 3 | 0 | 1 | Crus II, Dentate, Crus I, Lobule I_IV, Lobule VIIb |
| 3 | 0 | 2 | 0 | 3 | 3 | Crus II, Crus I, Lobule VIIb, Lobule I_IV, Lobule IX |
| 4 | 1 | 4 | 0 | 1 | 11 | Crus I, Crus II, Lobule VI, Lobule VIIb, Vermis VI |

Table 4: The distribution of white matter tracts for each parcel, and the top cerebellar regions connected by each parcel according to the SUIT parcellation [Diedrichsen et al., 2009; Diedrichsen et al., 2011]. Abbreviations of white matter tracts are: inferior cerebellar peduncle (ICP), middle cerebellar peduncle (MCP), superior cerebellar peduncle (SCP), Input and Purkinje (IP), parallel fibres (PF).

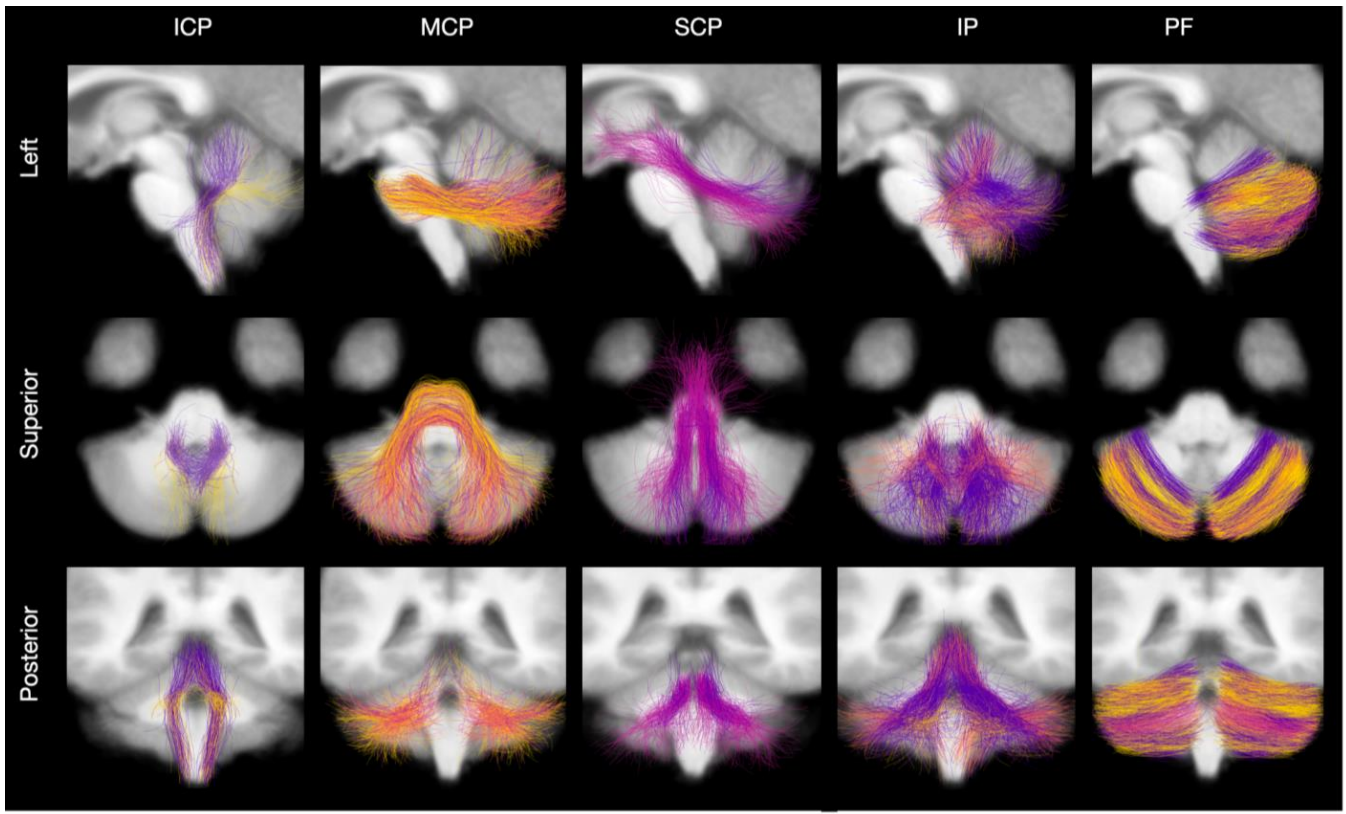

Fig 8: Visualisations of each white matter tract, where colour indicates parcel. White matter tracts shown are: inferior cerebellar peduncle (ICP), middle cerebellar peduncle (MCP), superior cerebellar peduncle (SCP), Input and Purkinje (IP), parallel fibres (PF)

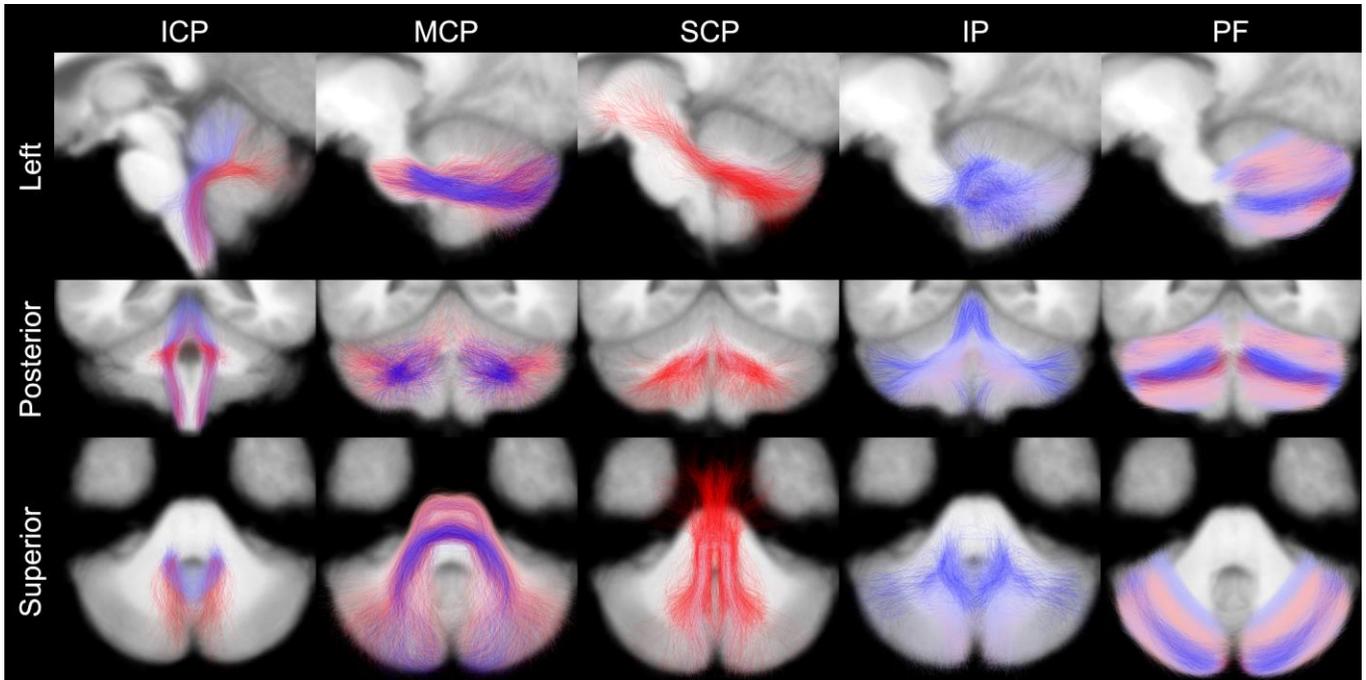

Fig 9: Parcel-specific tract visualisation of motor/cognitive distribution. Each fibre cluster for a given white matter tract is assigned the motor and cognitive mean saliency values across all fibre clusters that belong to the same tract and parcel. These motor and cognitive values are then combined into a single score by calculating: (motor - cognitive) / (motor + cognitive). These scores are visualised from blue (lower value = more cognitive) to red (higher value = more motor), with a white value indicating a middle neutral value of neither motor nor cognitive dominance. For visualisation purposes, the colours for each tract are normalised independently by dividing by the absolute maximum value across all fibre cluster scores within the tract.

## 4 Discussion

We have introduced Deep Multimodal Saliency Parcellation (DeepMSP), a novel deep learning approach for cerebellar pathway parcellation. We thoroughly experimented with all stages of the DeepMSP pipeline, including network selection, structure-function saliency representation, and clustering algorithm and cluster count. We found that the 1DCNN and transformer network architectures performed comparably for the multitask prediction of endurance, strength, reading decoding, and vocabulary comprehension, as well as both significantly outperforming a simpler fully connected network architecture for these same tasks. Investigating the saliencies of this trained transformer, which achieved slightly better mean performance on the evaluation metrics compared to the 1DCNN, we found that the minimum statistic for all microstructural measures (FA1, FA2, Trace1, Trace2) was more salient than commonly used statistics such as mean and variance. This finding is in line with other machine learning work that has found that using minima or maxima statistics for dMRI structural fibre cluster measures achieved higher classification performance than summary statistic measures such as the mean [Zhang et al., 2018a]. It is well known that the mean value of a microstructure measure (such as FA) within a fibre tract can be less informative than studying individual microstructure values along the tract [Colby et al., 2012; O'Donnell et al., 2009; Yeatman et al., 2012]. The minimum FA generally occurs at streamline endpoints, i.e. the white matter grey matter interface, an anatomical region that is highly predictive of individual cognitive performance [Chen et al., 2024]. In contrast, the minimum Trace value may occur in the central core (or "skeleton") of a fibre tract, another region that is considered highly informative for the study of the white matter [Smith et al., 2006]. Our findings suggest the importance of including summary statistics beyond the popular mean value for machine learning applications related to microstructure. In addition to this finding concerning microstructural measure statistics, we also observed that the majority of differences when comparing motor and cognitive saliencies were for clusters which had a high saliency when averaging over both categories. Finally, we found that a saliency representation that explicitly encoded category bias ("Category Displacements (3 elements)") resulted in better clustering performance, with a cluster count of 4 achieving the best internal cluster quality metric scores.

Applying DeepMSP, we identified four cerebellar pathway parcels that exhibited unique structure-function saliency patterns that were stable across training folds. We found that the parcel with the highest mean saliency was motor dominant, with a higher mean saliency for motor tasks compared to cognitive tasks. The opposite was true for the second most salient parcel, while the remaining 2 parcels had very similar saliencies for motor and cognitive tasks. However, the absolute differences in cognitive and motor saliencies were relatively small for each parcel. In the HCP-YA dataset, it can be observed that motor (strength/endurance) and cognitive (reading decoding/vocabulary comprehension) functional performance measures are correlated (average Pearson's $r$ = 0.19), making structure-function saliencies for motor and cognitive categories potentially difficult to disentangle. Hence, experimentation with categories beyond motor and cognitive function should be explored in future work. This parcellation also subdivided all cerebellar white matter tracts, where, for example, the 9 fibre clusters of the MCP were split across all four parcels. Each parcel also contained clusters from multiple white matter tracts, such as the most salient parcel that included clusters from the MCP, SCP and PF, and also connected multiple cerebellar regions, such as Crus II, Dentate, Crus I, Lobule I_IV, and Lobule VIIb. With the cerebellum playing a role in various disorders such as Parkinson's disease [Haghshomar et al., 2022], Alzheimer's disease [Toniolo et al., 2020], and autism spectrum disorder [Jeong et al., 2014], findings such as these, which have been produced via DeepMSP, may impact the diagnosis and treatment of such cerebellar-related disorders.

Several detailed observations about the experimental results are discussed below.

We demonstrated that the proposed 1DCNN and transformer models significantly outperformed a simpler fully connected model in terms of predictive ability, and we showed that the saliencies derived from the transformer were consistent across the five test-retest folds. We also showed that different parcels had unique motor and cognitive saliency patterns, as expected based on the known complex functionality of the human cerebellum. Finally, the resulting parcels demonstrated coherent spatial/anatomical patterns, even though no spatial or tract label information was input to the network. Overall, these findings support the potentially meaningful nature of the saliencies employed in this project, and the parcellations resulting from these saliencies. However, one aspect of network selection that remains for future research is to investigate the stability of the parcellations formed by training different architectures for the prediction of the same NIH Toolbox measures. If relatively consistent parcellations are produced across a variety of network architectures, it would further validate the DeepMSP pipeline, while also allowing for smaller and less computationally expensive networks to be used, increasing the accessibility of DeepMSP for those with limited computational resources. This is particularly relevant to the neuroimage analysis field, where simpler models have been shown to perform comparably to more complex state of the art models [Kugelman et al., 2022].

We focused our experimental results on the 4-parcel parcellation, which was the most successful according to the SC and DBI cluster quality metrics. These metrics are popularly used to evaluate the quality of tractography parcellation [Chen et al., 2023; Siless et al., 2013; Vázquez et al., 2020]. We relied on the use of cluster quality metrics for the selection of the parcel count, though it is well known that there may not be an optimal number of parcels in neuroimaging [Eickhoff et al., 2018]. This stems from the complex, multiscale organisation of the brain, where different applications that leverage a parcellation may require different levels of granularity [Eickhoff et al., 2018]. Recent work in parcellation of the cerebellar cortex suggests that a hierarchical, multiscale parcellation is effective to allow analysis at multiple granularities, where 4 parcels are chosen for the coarsest scale [Nettekoven et al., 2023]. Therefore, depending on the application of interest, a finer parcellation (more than 4 parcels) may be useful. For this reason, upon publication we will openly provide our 4-parcel parcellation, as well as parcellations of other scales on GitHub (https://github.com/SlicerDMRI/DeepMSP). We note that because the proposed parcellations are based on a finer fibre cluster parcellation that can be automatically applied across the lifespan [Zhang et al., 2018b] using open-source software (https://github.com/SlicerDMRI/whitematteranalysis), it is straightforward to apply the proposed parcellations for new studies.

Human brain parcellation is a crucial challenge [Arslan et al., 2018; Bijsterbosch et al., 2020] and is increasingly important for data reduction in large-scale neuroimaging studies [Cetin-Karayumak et al., 2023; Eickhoff et al., 2018]. There is a large body of work on cerebellar parcellation including cerebellar cortical parcellation, cerebellar deep nuclei segmentation, and atlasing of major cerebellar tracts [Buckner et al., 2011; Carass et al., 2018; Diedrichsen et al., 2009; Makris et al., 2005; Nettekoven et al., 2023]. We have aimed to situate our results in the context of this established work by using the well-known SUIT anatomical parcellation [Diedrichsen et al., 2009] and providing information about the cerebellar cortical connectivity of the proposed 4-parcel parcellation. In our study, the most salient parcel was primarily connected to Crus II and Crus I, which are considered to form part of key cognitive sociolinguistic and demand networks [Nettekoven et al., 2023], as well as the dentate nucleus, which participates in both motor and non-motor functions and is crucial in the communication between the cerebellum and the rest of the brain [Steele et al., 2017]. Our results generally suggest that motor and cognitive functions are distributed across the cerebellar pathways. This finding is in line with previous research that finds microstructural measures of the cerebellar white matter tracts are associated with motor and cognitive function and dysfunction [Cao et al., 2021; Chang et al., 2022; Chen et al., 2020; Fritz et al., 2022; Hernandez-Castillo et al., 2015; Hernandez-Castillo et al., 2016; Kim et al., 2021; Travis et al., 2015; Wu et al., 2017; Zekelman et al., 2023], as well as recent research identifying both motor and cognitive regions within the cerebellar dentate nucleus [Kulkarni et al., 2023; Palesi et al., 2021].

In addition, our results are supported by anatomical and physiological studies that show that cerebellar neurons, modules and regions receive sensory and motor signals from diverse sources (e.g., [Apps et al., 2018; Huang et al., 2013; Ishikawa et al., 2015; Schmahmann, 1996]). Our results also suggest that motor and cognitive saliencies are spatially organised across cerebellar cortical and subcortical white matter pathways. Indeed, anatomical studies have found that many sensory, motor, and association regions of the cerebral cortex communicate with the cerebellum via the pontine nuclei, which indicates that the signals carried in the middle cerebellar peduncle are functionally diverse (e.g., [Kelly and Strick, 2003; Schmahmann et al., 2004; Schmahmann and Pandya, 1997a; Schmahmann and Pandya, 1997b]). Similarly, the inferior cerebellar peduncle carries sensory signals from the spinal cord, motor signals to vestibular nuclei, and cognitive information likely through the principal nuclei of the inferior olivary nuclear complex (e.g., [Koziol et al., 2014; Schmahmann, 2010]). Overall our findings are in line with the notion that the functional organisations of the brain extend beyond the cortical surface and are reflected in the composition of white matter [Ghimire et al., 2021]. Interrelating the diverse signals in cerebellar white matter with functional localisation in the cerebellar cortex and deep cerebellar nuclei is an important next step that this method may begin to address [Buckner et al., 2011; Makris et al., 2003; Makris et al., 2005; Stoodley and Schmahmann, 2009; Stoodley and Schmahmann, 2018]. Looking forward, with the cerebellar pathways playing a role in various disorders such as Parkinson's disease [Haghshomar et al., 2022], Alzheimer's disease [Toniolo et al., 2020], and autism spectrum disorder [Jeong et al., 2014], future work in parcellation of cerebellar pathways may impact the study of such cerebellar-related disorders.

As this is a completely novel approach to cerebellar pathway parcellation, there are a few current limitations and potential directions for future work. In regard to the deep learning and saliency computation stages of DeepMSP, we focused on one high-performing network (multitask transformer) designed to provide saliencies for the prediction of multiple motor and cognitive individual functional performance measures. However, a different network could potentially focus on a different aspect of the high-dimensional input data, producing different saliency patterns. We believe we have somewhat ameliorated this concern by averaging the saliency across many dMRI features and across over 1,000 testing subjects to base our final parcellation on robust input saliency features representative of a large population. However, the investigation of different saliency measures (e.g. LIME [Ribeiro et al., 2016], Guided Backpropagation [Springenberg et al., 2014], Guided Grad-CAM [Selvaraju et al., 2020]) and the effect of different networks is of interest for future research, especially when considering additional individual functional performance measures that may be of interest to inform parcellation for future studies.

We also believe it may be of interest for future work to investigate measures of saliency that incorporate directional or other forms of information beyond purely the magnitude of predictive importance. Although we have successfully shown that saliency varies within and across anatomical tracts, a parcel containing fibre clusters of similar saliency suggests that these fibre clusters have similar predictive importance, and not necessarily that they have a similar relationship with an individual functional performance measure of interest. For example, a fibre cluster where FA is positively correlated with an individual functional performance measure, and another where it is negatively correlated, and yet another with a nonlinear relationship could be considered similarly important (i.e. similar saliencies) by a deep learning model. However, as described in section 2.5, we have exclusively focused on the magnitude of the saliency values, while disregarding the value's sign. Future work could investigate this aspect more deeply.

Additionally, the proposed parcellation is based on a particular tractography algorithm (which is consistent across the lifespan and robust) that provides a particular set of fibre-specific microstructure measures that are estimated during fibre tracking based on a multi-tensor model [Malcolm et al., 2010; Reddy and Rathi, 2016; Zhang et al., 2018b]. While the proposed parcellation can be applied straightforwardly to tractography computed using other models (via application of our fibre cluster atlas [Zhang et al., 2018b]), it is not known if microstructure measures from another model would have similar saliency patterns to those observed here. Additionally, we have investigated bilateral parcellation, a popular strategy for cerebellum parcellation [Nettekoven et al., 2023]. However, as we did not explore an asymmetric approach to structure-function parcellation, the degree to which the structure-function saliency patterns are asymmetric remains unknown. Hence future work may benefit from exploring a lateralized approach.

Finally, while the NIH Toolbox assessments utilised in this study are an established and standard approach to measuring cognitive and motor function, these assessments measure limited and broad aspects of cognitive behaviour and motor performance [Hodes et al., 2013; Reuben et al., 2013; Weintraub et al., 2013]. We anticipate that more refined and controlled outcome measures may reveal a stronger relationship of sensory, motor, and cognitive functions with fibre clusters in the cerebellum. Additionally, the tractography and fibre-cluster-based quantitative measurements in this study were performed using high-quality HCP-YA data. However, it is important to note that the fine-grained anatomy of the cerebellum requires higher resolution scanning to depict, for example, the fine folds of the cerebellar cortical folia [Sereno et al., 2020]. Partial volume effects in the current study were ameliorated by using measurements (e.g. FA or NoS) performed at the level of a fine fibre cluster parcellation, where each cluster occupied a relatively small region of white or grey matter, and where fibre-specific tissue microstructure measures were computed from a multi-tensor model. However, future work using higher resolution diffusion MRI [Ramos-Llordén et al., 2020] may enable more detailed insights into finer scale cerebellar pathway parcellation. We also note the fibre clusters in the employed tractography atlas are biased towards the longer connections in the posterior lobe of the cerebellum, the neocerebellum. In this way, our investigation may be biased towards phylogenetically newer functions of the cerebellum. Thus, future investigations may continue to investigate across the entire cerebellum.

Looking forward, there are many future research directions for building on DeepMSP. Firstly, the quality of the saliency values used in DeepMSP depends on both the prediction performance of the deep network, and the saliency calculation algorithm. Hence, it could be beneficial to experiment with more architectures, larger datasets or augmentation strategies, and other well-known saliency computation methods such as LIME [Ribeiro et al., 2016], Guided Backpropagation [Springenberg et al., 2014], and Guided Grad-CAM [Selvaraju et al., 2020]. Secondly, our saliencies encoded the magnitude of the predictive importance of a particular fibre cluster, hence future work could explore directional structure-function relationships.

# 5 Conclusion

In this work, we proposed DeepMSP, a novel saliency-based parcellation framework for multimodal, data-driven tractography parcellation. Our method generates parcellations that aim to reflect the importance of pathway microstructure and connectivity for predicting individual cognitive and motor functional performance measures. Through utilising both structural features and functional performance measures, this parcellation strategy may have the potential to enhance the study of structure-function relationships of the cerebellar pathways.

# Supplementary Material

| Metric | Description | Equation |
|---|---|---|
| Silhouette coefficient (SC) [Rousseeuw, 1987] | Compares the mean within-cluster distance for a given point, to the point's mean nearest-neighbour cluster distance. We report the mean SC, which is a single value representing the quality of the entire clustering, ranging from -1 (worst) to 1 (best). | $$SC = \frac{d_2 - d_1}{\max(d_1, d_2)}$$ where $d_1$ is the distance of a datapoint to all other datapoints in its cluster, and $d_2$ is the distance between a datapoint and all other datapoints in the nearest neighbour cluster. |
| Davies-Bouldin index (DBI) [Davies and Bouldin, 1979] | The average similarity between each cluster and its most similar cluster, where similarity is a function of cluster diameter and distance between clusters. A lower value for this metric is considered better, with 0 being a lower bound. For the complete equations for these metrics, see supplementary materials. | $$DBI = \frac{1}{k} \sum_{i=1}^{k} \max_{i \neq j}(\text{sim}_{ij})$$ $$\text{sim}_{ij} = \frac{c_i + c_j}{d_{ij}}$$ where $c_n$ is the mean distance of each point in cluster n to its centroid, and $d_{ij}$ is the distance between the centroids of clusters i and j. |

Table S1: Descriptions and equations for the standard intrinsic cluster quality metrics we used.

| Category | NIH Toolbox Measure | Fully Connected | 1DCNN | Transformer |
|---|---|---|---|---|
| Motor | Endurance | 0.09 (± 0.07) | 0.23 (± 0.06) | 0.25 (± 0.02) |
| | GaitSpeed | 0.04 (± 0.06) | 0.01 (± 0.13) | -0.02 (± 0.11) |
| | Dexterity | 0.03 (± 0.09) | 0.14 (± 0.06) | 0.08 (± 0.04) |
| | Strength | 0.22 (± 0.17) | 0.57 (± 0.04) | 0.57 (± 0.05) |
| Cognitive | PicSeq | 0.04 (± 0.10) | 0.09 (± 0.04) | 0.11 (± 0.04) |
| | CardSort | 0.04 (± 0.10) | 0.08 (± 0.05) | 0.12 (± 0.07) |
| | Flanker | 0.02 (± 0.04) | 0.10 (± 0.01) | 0.14 (± 0.06) |
| | ReadEng | 0.02 (± 0.07) | 0.20 (± 0.05) | 0.21 (± 0.05) |
| | PicVocab | 0.02 (± 0.06) | 0.21 (± 0.07) | 0.25 (± 0.07) |
| | ProcSpeed | 0.04 (± 0.06) | 0.08 (± 0.06) | 0.07 (± 0.07) |
| | ListSort | 0.05 (± 0.08) | 0.12 (± 0.14) | 0.14 (± 0.06) |

Table S2. The Pearson correlation coefficients of the fully connected, 1DCNN, and transformer models for predicting all 11 NIH Toolbox measures. Values are the mean across the 5 folds of cross-validation, with the standard deviation indicated in brackets.

| Category | NIH Toolbox Measure | Fully Connected | 1DCNN | Transformer |
|---|---|---|---|---|
| Motor | Endurance | 17.89 (± 0.41) | 11.24 (± 0.76) | 11.34 (± 0.76) |
| Motor | GaitSpeed | 1.56 (± 1.24) | 0.68 (± 0.30) | 0.90 (± 0.54) |
| Motor | Dexterity | 15.62 (± 1.06) | 8.29 (± 0.34) | 8.28 (± 0.26) |
| Motor | Strength | 19.88 (± 1.36) | 14.51 (± 1.34) | 13.51 (± 1.25) |
| Cognitive | PicSeq | 19.24 (± 1.19) | 14.25 (± 0.95) | 13.72 (± 0.70) |
| Cognitive | CardSort | 15.97 (± 0.62) | 9.11 (± 0.77) | 8.44 (± 0.72) |
| Cognitive | Flanker | 15.71 (± 0.60) | 9.29 (± 1.53) | 8.55 (± 0.32) |
| Cognitive | ReadEng | 18.64 (± 0.89) | 12.49 (± 1.01) | 12.38 (± 0.92) |
| Cognitive | PicVocab | 18.70 (± 0.55) | 12.01 (± 0.75) | 11.81 (± 0.94) |
| Cognitive | ProcSpeed | 21.31 (± 1.11) | 17.41 (± 1.09) | 16.57 (± 1.21) |
| Cognitive | ListSort | 17.27 (± 0.44) | 10.72 (± 0.63) | 11.00 (± 0.33) |

Table S3. The mean absolute errors of the fully connected, 1DCNN, and transformer models for predicting all 11 NIH Toolbox measures. Values are the mean across the 5 folds of cross-validation, with the standard deviation indicated in brackets.

| Category | NIH Toolbox Measure | Pearson's r | | | | MAE | | | |
|---|---|---|---|---|---|---|---|---|---|
| | | ANOVA | Post-hoc Tukey test p-values | | | KW | Post-hoc Dunn's test p-values | | |
| | | | FC vs 1D | FC vs. TR | 1D vs. TR | | FC vs 1D | FC vs. TR | 1D vs. TR |
| Motor | Endurance | 0.003 | 0.005 | 0.001 | **0.762** | 0.008 | 0.011 | 0.049 | **1.000** |
| Motor | Strength | 0.002 | 0.001 | 0.001 | **0.998** | 0.007 | **0.071** | 0.007 | **1.000** |
| Cognitive | ReadEng | 0.003 | 0.001 | 0.001 | **0.988** | 0.009 | 0.033 | 0.017 | **1.000** |
| Cognitive | PicVocab | 0.003 | 0.002 | 0.001 | **0.748** | 0.009 | 0.027 | 0.022 | **1.000** |

Table S4. P-values for all statistical significance testing. Comparisons that were not statistically significantly different (p >= 0.05) are bolded. Abbreviations are as follows: FC = fully connected network, 1D = 1DCNN, TR = Transformer, KW = Kruskal-Wallis test.

Saliency-based Cerebellar Parcellation

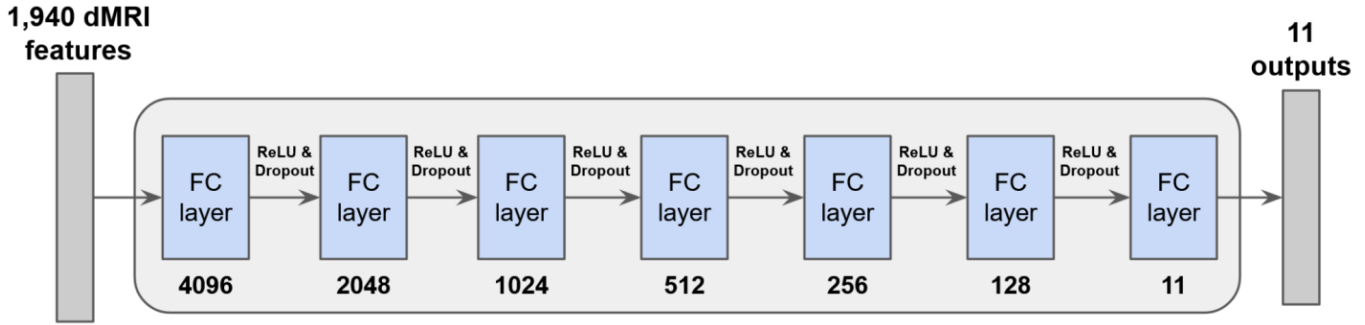

Fig S1: Architecture diagram of our fully connected network. 'FC layer' indicates a fully connected layer, with the output size being indicated under each box.

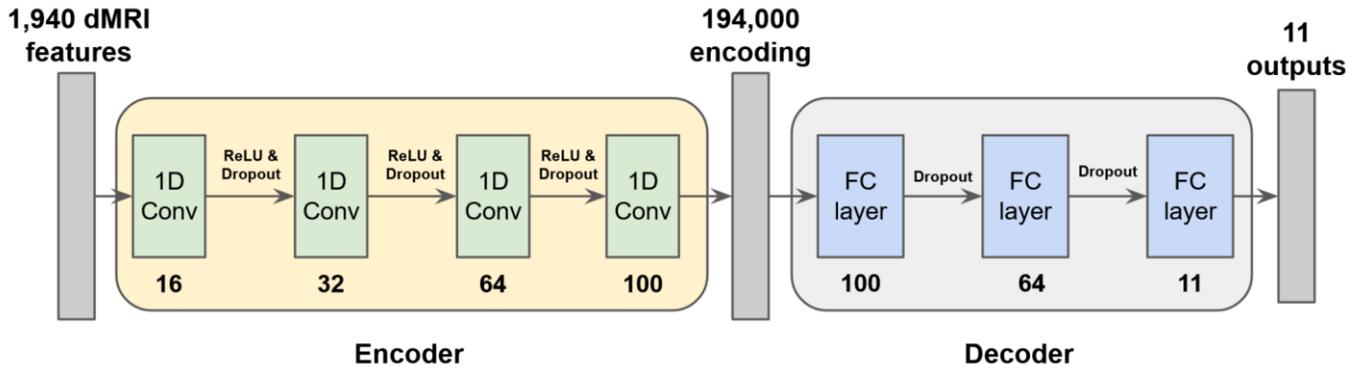

Fig S2: Architecture diagram of our 1DCNN. '1D Conv' indicates 1D-convolution with a kernel size of 5, stride 1, padding 2, and the number of filters indicated in text under the box. 'FC layer' indicates a fully connected layer, with the number underneath indicating the output size.

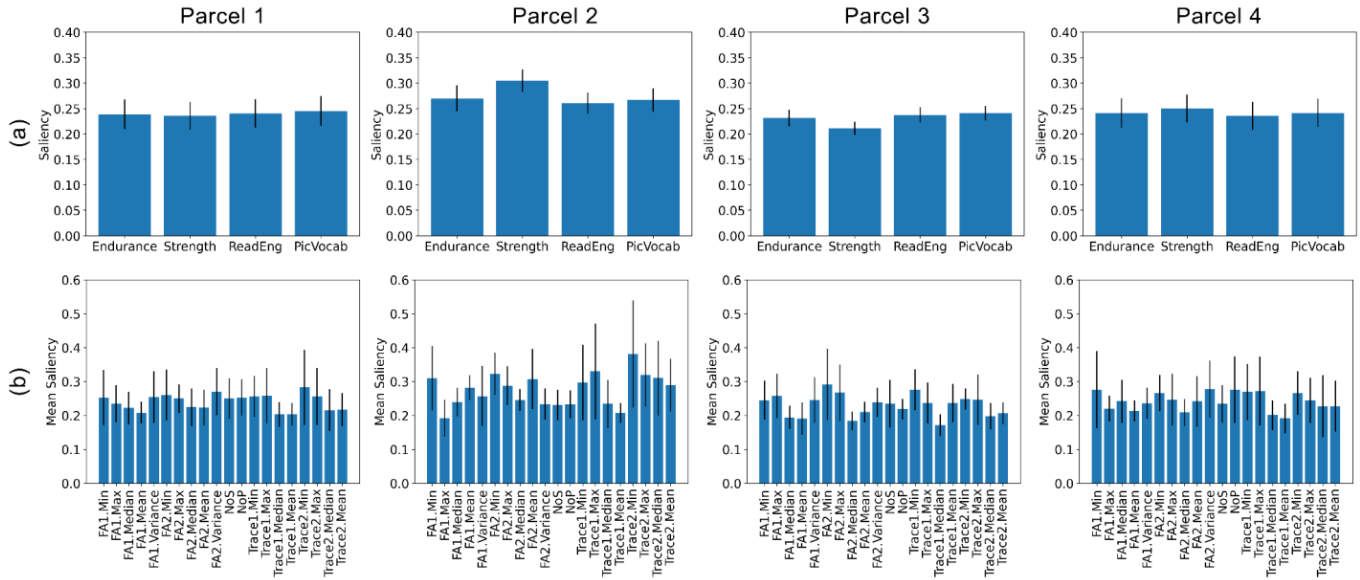

Fig S3: The mean saliency for each parcel for (a) NIH Toolbox measures, (b) dMRI features. Bar height indicates the mean saliency across all fibre clusters within the parcel, with error bars indicating standard deviation.